\documentclass[prb,letterpaper,aps,floatfix,twocolumn]{revtex4-2}

\usepackage{graphicx}
\usepackage{amsmath,amssymb}

\usepackage[pdfstartview=FitH,breaklinks=true,bookmarks=true,colorlinks=true,anchorcolor=black,citecolor=red,filecolor=black,menucolor=black,urlcolor=blue,linkcolor=blue]{hyperref}

\begin{document}

\title{Unsupervised learning of phase transitions via modified anomaly detection with autoencoders}

\author{Kwai-Kong Ng}
\email[]{kkng@thu.edu.tw}
\affiliation{Department of Applied Physics, Tunghai University, Taichung 40704, Taiwan}
\author{Min-Fong Yang}
\email{mfyang@thu.edu.tw}
\affiliation{Department of Applied Physics, Tunghai University, Taichung 40704, Taiwan}

\date{\today}

\begin{abstract}
In this paper, a modified method of anomaly detection using convolutional autoencoders is employed to predict phase transitions in several statistical mechanical models on a square lattice. We show that, when the autoencoder is trained with input data of various phases, the mean-square-error loss function can serve as a measure of disorder, and its standard deviation becomes an excellent indicator of critical points. We find that various types of phase transition points, including first-order, second-order, and topological ones, can be faithfully detected by the peaks in the standard deviation of the loss function. Besides, the values of transition points can be accurately determined under the analysis of finite-size scaling. Our results demonstrate that the present approach has general application in identification/classification of phase transitions even without a priori knowledge of the systems in question.

\end{abstract}

\maketitle

%%%%%%%%%%%%%%%%%%%%%%%%%%%%%%%%%%%%%%%%%%%%%%%%%%%%%%%%%%%%%%%%%%
\section{INTRODUCTION} \label{sec:intro}

Identifying phases of matter and their transitions is an essential research focus in the areas of statistical and condensed-matter physics. Usually, relevant order parameters or correlation functions are measured to classify different phases of matter. Nonetheless, this approach demands detailed knowledge of systems and thus becomes challenging to be applied to systems where no conventional order parameter exists, such as spin liquids and materials with topological properties.

Recent developments in machine learning have opened new avenues to process and find correlations in complex data. This inspires various proposals of applying machine learning techniques in detecting phase transitions directly from synthetic data~\cite{melkonat1,spinlic2,Zhang-Liu-Wei2019,%review1
Ho_2021,Ho-Wang2023,Chertenkov_etal2023,%
Wang2016,huber1,Wetzel2017,Wang_etal2021,Miyajima_2021,%
Tasi_etal2021,Giataganas:2021jqm,Chung_etal2023,ng2022BKT}. %
The advantage of machine learning in the classification of phases is that it can analyze large datasets quickly and accurately. This helps in identifying patterns or changes in data that indicate a phase transition, which can be difficult to be recognized by using traditional methods.
These proposals can be differentiated by the degree of a priori knowledge required.
In order to discover unknown phases in complex systems, the method of choice may be the so-called unsupervised machine learning~\cite{Wang2016,huber1,Wetzel2017,Wang_etal2021,Miyajima_2021,%
Tasi_etal2021,Giataganas:2021jqm,Chung_etal2023,ng2022BKT}, %
which requires no prior labeling on a dataset and builds knowledge directly from analyzing the data structure.

Recently, an automated and unsupervised machine learning based on anomaly detection is put forward to find regions of interest for possible new phases~\cite{Kottmann_etal2020,Kottmann_etal2021,Acevedo_etal2021,Kaming_2021}. The main steps along this approach are given as follows. One first trains an autoencoder to reproduce a certain class of data until the loss function given by the mean squared error (MSE) becomes small enough. The transition to a different class of data can then be discriminated by monitoring the abrupt change of MSE which signals the anomaly. This approach is particularly useful to reveal subtle features of a condensed matter system which can remain hidden from any conventional regression or classification scheme. Notably, this method works well even by taking quantities that arise naturally from the state description without further processing as input data. Hence the necessity for defining and calculating suitable observables to identify the phases can be circumvented. While the phase diagrams can be successfully mapped out, it is unclear whether this method can produce quantitatively reliable values of phase boundaries.

In this work, we revisit this unsupervised method for phase characterization. 
We propose a modified method of anomaly detection in which, instead of using one particular class of data, we train the antoencoder with all class of data within the region of interest. In this scheme, the MSE behaves as a "disorder" parameter among different phases, and the standard deviations of MSE reveal distinct peaks at the boundaries of different phases.    
Our findings highlight that the standard deviation of MSE should be a better indicator for phase transitions compared to MSE alone. Notably, our approach is robust and universal, as it does not rely on prior knowledge of the number of phases or the locations of ordered or disordered phases within the phase diagram, making it applicable to a wide range of statistical models.
 
For illustration, we explore phase transitions in several classical spin models on a square lattice, including the $q$-state Potts model, the $q$-state clock model, and the generalized XY model. We find that the peaks in the standard deviation of MSE always give the phase transitions faithfully, no matter whether the transitions are of first-order, second-order, or Berezinskii-Kosterlitz-Thouless (BKT) types. In addition, good agreement of the transition points with previous findings in the literature can be achieved after the finite-size scaling analysis. This shows that the modified anomaly detection by analyzing the standard deviation of MSE should be a useful approach among other unsupervised methods in identification/classification of phase transitions.

The remainder of this paper is organized as follows.
The general approach of applying modified anomaly detection with neural networks to map out phase diagrams is described in section~\ref{sec:method}.
Our results for the $q$-state Potts model, the $q$-state clock model, and the generalized XY model are presented in \ref{sec:q_Potts}, \ref{sec:q_clock}, and \ref{sec:GXYmodel}, respectively.
We conclude our paper in section~\ref{sec:conclusion}.

%%%%%%%%%%%%%%%%%%%%%%%%%%%%%%%%%%%%%%%%%%%%%%%%%%%%%%%%%%%%%%%%%%
\section{modified anomaly detection method} \label{sec:method}

\begin{figure}[htb]
\includegraphics[width=\linewidth]{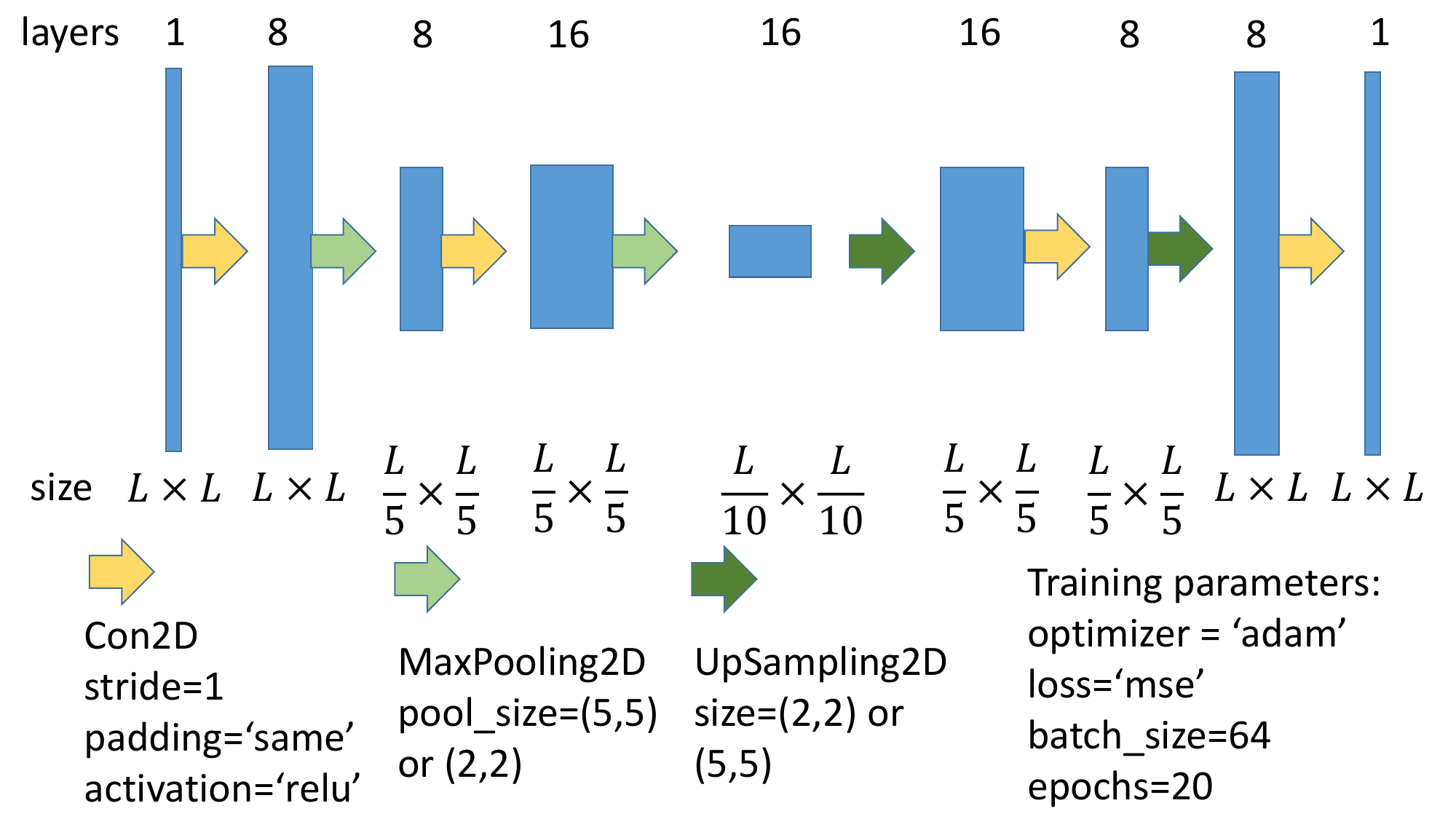}
\caption{Schematic representation of an autoencoder. The input data is compressed by the encoder and the decoder expands the compressed data to its original size. The intermediate space with compressed dimension ($\frac{L}{10} \times \frac{L}{10}$) is called the latent space. As discussed in Secs.~\ref{sec:q_clock} and~\ref{sec:GXYmodel}, for the $q$-state clock and the generalized XY models, the input/output layers consist of two and four layers, respectively. The hyperparamters of the neural network is also presented.}
\label{fig:CAE}
\end{figure}

In our procedure of modified anomaly detection, we utilize a conventional convolutional autoencoder (CAE) architecture~\cite{baldi2012autoencoders,makhzani2015adversarial,vincent2010stacked}, which constitutes a deep neural network structure as illustrated in Fig.~\ref{fig:CAE}. The CAE serves as a multilayer neural network designed to achieve dimensionality reduction. It consists of two essential components: an encoder ($e$) and a decoder ($d$), typically arranged symmetrically and trained jointly to minimize the reconstruction error associated with the data. The encoder learns a non-linear transformation function $e:\mathcal{X}\to\mathcal{Z}$, which projects the input data from its original high-dimensional space $\mathcal{X}\equiv\lbrace x\rbrace$ to a lower-dimensional latent space $\mathcal{Z}\equiv\lbrace z\rbrace$ lying in a hidden layer. The hidden layer locates between the encoder and the decoder and its dimension controls the dimensionality of the reduced data representation. In the present study, we focus on classical spin models and the input data corresponds to a set of spin configurations of size $L\times L$. In order to reach better results, we opt for a reduced dimension of $\frac{L}{10}\times\frac{L}{10}$ for the latent space, instead of an $L$-independent size employed usually. The decoder, in turn, learns a non-linear transformation $d:\mathcal{Z}\to\mathcal{X}$ that maps the latent vectors $z=e(x)$ back to the original high-dimensional input space $\mathcal{X}$. Consequently, the latent vector $z=e(x)$ is transformed in order to reconstruct the original input data, yielding $\hat{x}=d(z)=d\left(e(x)\right)$, where $\hat{x}$ represents the output for a given input $x$. The optimization of the CAE involves minimizing the reconstruction error between the input $x$ and the output $\hat{x}$ with respect to the training data. The reconstruction error is quantified through the mean square error (MSE):
\begin{equation}\label{eqn:MSE}
\text{MSE}=\frac{1}{n L^2}\sum_{s=1}^{n} \sum_{i=1}^{L^2} |x_i^s-\hat{x}_i^s|^2 \; ,
\end{equation}
where $n$ is the number of data in the training or testing dataset.

Prior studies have demonstrated the utility of the MSE of autoencoders in discerning phase boundaries within various classical and quantum models~\cite{Kottmann_etal2020,Kottmann_etal2021,Acevedo_etal2021,Kaming_2021}. In these methodologies, a designated region of the phase diagram is chosen to represent normal data and is subsequently tested across the entire diagram. An anomaly, characterized by significant increases or decreases in MSE, becomes evident when testing states belong to the phases other than the trained one. Between those in the training region and these states lies a transition from normal to anomalous data, corresponding to a phase transition. Consequently, the points at which the MSE changes abruptly are commonly regarded as phase boundaries.
As noted in Ref.~\cite{Kottmann_etal2021}, such a method of anomaly detection bears similarities to the fidelity approach~\cite{Zanardi_etal2007,Zhou_etal2008,Gu2010}, in which a phase transition is determined from the drop in the overlap (fidelity) between neighboring ground states in the phase diagram. To map out the complete phase diagram by this anomaly detection approach, it is recommended to iteratively train the model on anomalous regions until no new anomalous region is found. Therefore, several training processes proportional to the number of phases in systems are needed.

%------------------------- figure -------------------------------------
%------------------------- figure -------------------------------------

In our current work, in contrast to prior studies, we extract training data from various phases, enabling the CAE to be trained to minimize the MSE for all available phases within the dataset. 
As we will elaborate in the next section, in this framework, the MSE behaves as a measure of the degree of disorder across all phases within the training data. Consequently there is no need to select some particular classes of data for separate trainings if the dataset contains multiple phases. One notable advantage of our modified anomaly detection approach is efficiency, as it requires only a single training session to identify all phases within the phase diagram.
The dataset is generated using the classical Monte Carlo method with the Wolff algorithm, specifically chosen to mitigate critical slowing down near transition temperatures. For each temperature $T_i$, we prepare 1000 sets of independent spin configurations. To investigate the temperature range [$T_0$, $T_1$], we select ten temperatures evenly spaced within this range as the training set, with 20$\%$ reserved for validation purposes. This training set, encompassing configurations from multiple phases, is subsequently employed for training the CAE.
Note that the input spin values are always normalized within the range from 0 to 1. Besides, for the $q$-state Potts model, while the input spin values are discrete, we do not impose any constraints on the reconstructed output spins. Therefore, the reconstructed output values can take any real numbers within the range from 0 to 1.
Detailed information regarding the structure and hyperparameters of the CAE model is provided in Fig.~\ref{fig:CAE}. It is essential to emphasize that the CAE is optimized by minimizing the reconstruction errors associated with all phases, rather than focusing solely on one specific phase. The trained model is subsequently applied to all other temperatures, resulting in the MSE values, which represent the average square error of 1000 configurations for each temperature [Eq.~\eqref{eqn:MSE}].

Instead of relying on the sudden change in the MSE, we find that the standard deviation of the MSE, denoted as $\Delta_\mathrm{MSE}$, provides a more robust indicator for identifying the phase transitions. Similar proposal has been put forward in some previous investigations but in other context~\cite{Ho-Wang2023,Chertenkov_etal2023}. This may not be surprising since that thermal/quantum fluctuations are greatly enhanced around the transition points and then induce significant increase in the standard deviation of the MSE.
In the present study, the rescaled standard deviation $L\Delta_\mathrm{MSE}$,  instead of $\Delta_\mathrm{MSE}$ itself, is measured and we find that the peaks of $L\Delta_\mathrm{MSE}$ faithfully indicate the phase transition points. Moreover, after finite-size scaling, the extrapolated values of transition points agree well with either the exact ones (if available) or the best available results in the literature. 
Our investigation thus shows that the modified anomaly detection with CAE can be a quantitatively reliable and universal way to identify phase transitions as long as one focuses on the rescaled standard deviations.

The inclusion of a factor of $L$ in $\Delta_\mathrm{MSE}$ stems from the proportional scaling of all hidden layers within our CAE with respect to the linear size $L$. To illustrate, if the input linear size $L$ is increased by a factor of $M$ to obtain $L'=ML$, then sizes of all layers are also enlarged by a factor of $M$. This enlarged CAE can be approximated as a composite of $M^2$ identical CAEs of the original size $L^2$. Consequently, as the number of data points used in computing the MSE increases by $M^2$ times, the standard deviation $\Delta_\mathrm{MSE}$ is concurrently reduced by a factor of $1/M$. However, the rescaled standard deviation receives no such naive size dependence, $L'\Delta'_\mathrm{MSE}=(ML)\cdot(\Delta_\mathrm{MSE}/M)=L\Delta_\mathrm{MSE}$. Notably, away from the critical temperature, this rescaled standard deviation is nearly identical for all different sizes, as demonstrated in our calculations [see, for example, Fig.~\ref{fig:q2_potts}(b)].

In the following sections, we apply the aforementioned approach to investigate the $q$-state Potts model, the $q$-state clock model, and the generalized XY model. Through anomaly detection, we successfully identify all distinct phases for each model, including the topological BKT phases in the clock model and the nematic phases in the generalized XY model. By conducting finite-size analysis of $L\Delta_\mathrm{MSE}$, we can accurately determine the critical temperatures for all cases. Furthermore, by examining the scaling behavior of the MSEs, we can even distinguish between the first-order phase transitions and the second-order ones. The results for each model are presented in the subsequent sections accordingly.

%%%%%%%%%%%%%%%%%%%%%%%%%%%%%%%%%%%%%%%%%%%%%%%%%%%%%%%%%%%%%%%%%%
\section{$q$-state Potts model} \label{sec:q_Potts}

We first consider the $q$-state Potts model on a square lattice for the demonstration. It is found that our strategy is not only able to pinpoint the critical temperatures $T_c$'s but also to classify different types of phase transitions.

The $q$-state Potts model is a generalization of Ising model with rich contents and offers agents to study ferromagnet and certain other physics of solid states~\cite{Potts,Potts_review}. Its Hamiltonian reads
\begin{equation}\label{eqn:Potts model}
H_\mathrm{Potts} = -\sum_{\langle i,j\rangle} \delta(s_i,s_j)\;,
\end{equation}
where $\langle i,j\rangle$ denotes nearest neighbor sites, $\delta$ is the Kronecker delta function, and $s_i=n_i/(q-1)$ with the state index $n_i\in\{0,1,\cdots,q-1\}$ at the $i$-th site. The Ising model corresponds to the case of $q=2$.
The phase transition is known to be second-order for $1<q\leq 4$ and first-order for $q>4$ with the critical temperature $T_c=1/\ln(1+\sqrt{q})$~\cite{Potts,Potts_review}.

%------------------------- figure -------------------------------------
\begin{figure}[htb]
\includegraphics[width=\linewidth]{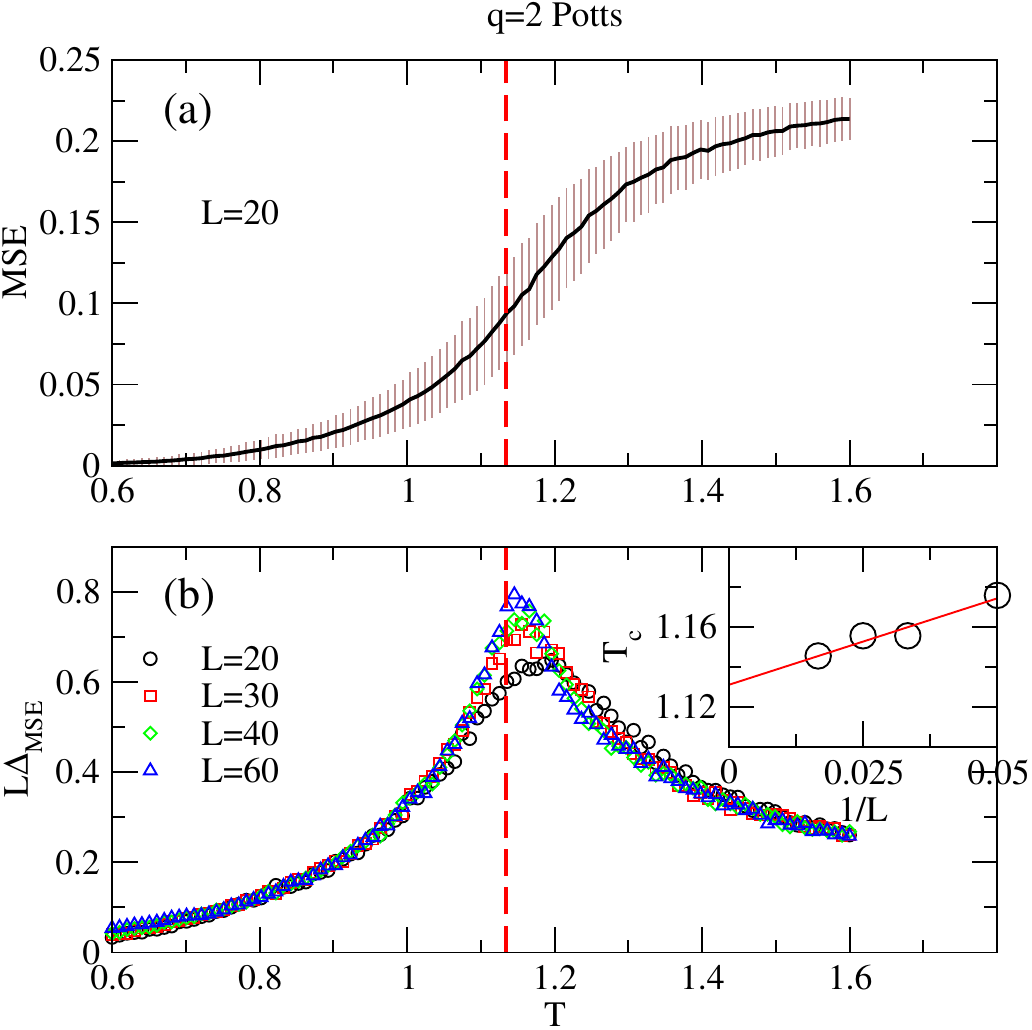}
\caption{(a) Mean square error MSE and (b) rescaled standard deviation $L\Delta_\mathrm{MSE}$ as functions of temperature $T$ for $q=2$ Potts model. The red dashed lines show the theoretical value of critical temperature $T_c=1/\ln(1+\sqrt{2})\cong1.135$. The error bars in (a) correspond to the standard derivation $\Delta_\mathrm{MSE}$. For the sake of clarity, only the data of linear size $L=20$ is shown. The inset in (b) shows the finite-size analysis of the critical temperature $T_c$ for different sizes, which gives the extrapolated value $T_c=1.131$.
}
\label{fig:q2_potts}
\end{figure}
%------------------------- figure -------------------------------------
%
%------------------------- figure -------------------------------------
\begin{figure}[htb]
\includegraphics[width=\linewidth]{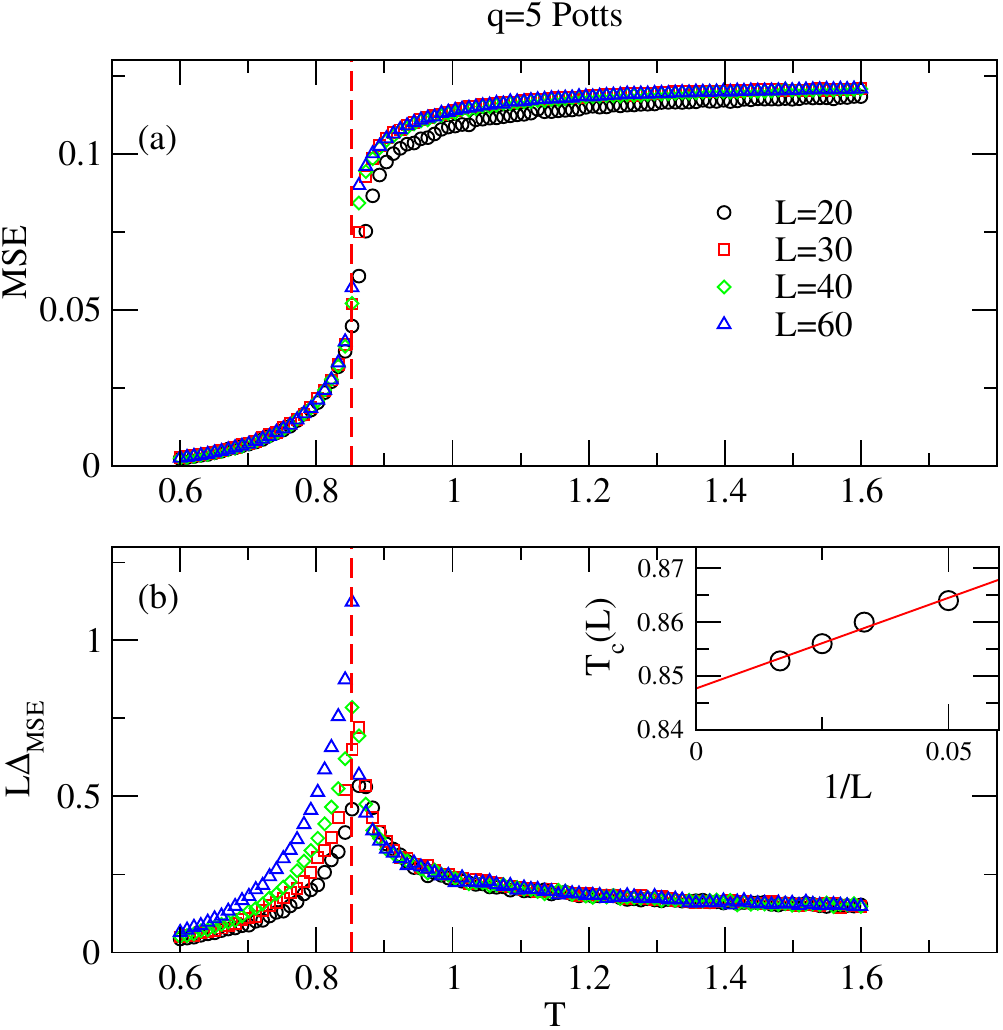}
\caption{(a) Mean square error MSE and (b) rescaled standard deviation $L\Delta_\mathrm{MSE}$ for several sizes as functions of temperature $T$ for $q=5$ Potts model. The red dashed lines show the theoretical value of critical temperature $T_c=1/\ln(1+\sqrt{5})\cong0.852$. The inset in (b) shows the finite-size analysis for the critical temperature $T_c$, which gives the extrapolated value $T_c=0.848$.
}
\label{fig:q5_potts}
\end{figure}
%------------------------- figure -------------------------------------

Following the procedure described in Sec.~\ref{sec:method}, we present our  results of the MSE and the rescaled standard deviation $L\Delta_\mathrm{MSE}$ for the cases of $q=2$ and $q=5$ in Figs.~\ref{fig:q2_potts} and \ref{fig:q5_potts}, respectively. We find that the value of MSE always increases monotonically from the low-temperature ordered phase to the high-temperature disordered one and thus can be viewed as a measure of disorder. This can be understood as follows.
Recall that the MSE of a state quantifies the CAE's ability to faithfully reconstruct the input state. Therefore, in the case of ordered input states with simple structures, such as ferromagnetic states with $s_i$ being site-independent, the CAE can effectively extract most of the information from the input state and reconstruct it with minimal loss, resulting in a very small value of MSE. Conversely, in situations of completely disordered states with random spin values, the CAE fails to capture specific information about the input state and typically produces an output state with an approximate average spin value, $(1/q)\sum_{n_i}s_i=1/2$, on each site. This gives the maximum expected value of MSE, $(1/q)\sum_{n_i}(s_i-1/2)^2=(q+1)/[12(q-1)]$, in the high-temperature limit, as shown in Figs.~\ref{fig:q2_potts}(a) and \ref{fig:q5_potts}(a).

While the MSE behaves as a measure of the degree of disorder, as illustrated in Fig.~\ref{fig:q2_potts}(a), it may display a gradual change even in the vicinity of an order-disorder phase transition. This makes it potentially inconvenient for precise determination of transition points. We notice that the (rescaled) standard deviation can be regarded as the corresponding ``susceptibility" of the MSE and thus should exhibit a pronounced peak at the transition point due to significant thermal fluctuations therein.
 This expectation is confirmed by our data. As seen from Figs.~\ref{fig:q2_potts}(b) and \ref{fig:q5_potts}(b), the rescaled standard deviation $L\Delta_\mathrm{MSE}$ exhibits a pronounced peak as the system size $L$ increases, particularly near the theoretical critical temperature $T_c$ indicated by the dashed line. Upon extrapolation to the thermodynamic limit as $L\rightarrow\infty$, the obtained values for $T_c$ are in excellent agreement with the theoretical values.

%------------------------- figure -------------------------------------
\begin{figure}[htb]
\includegraphics[width=\linewidth]{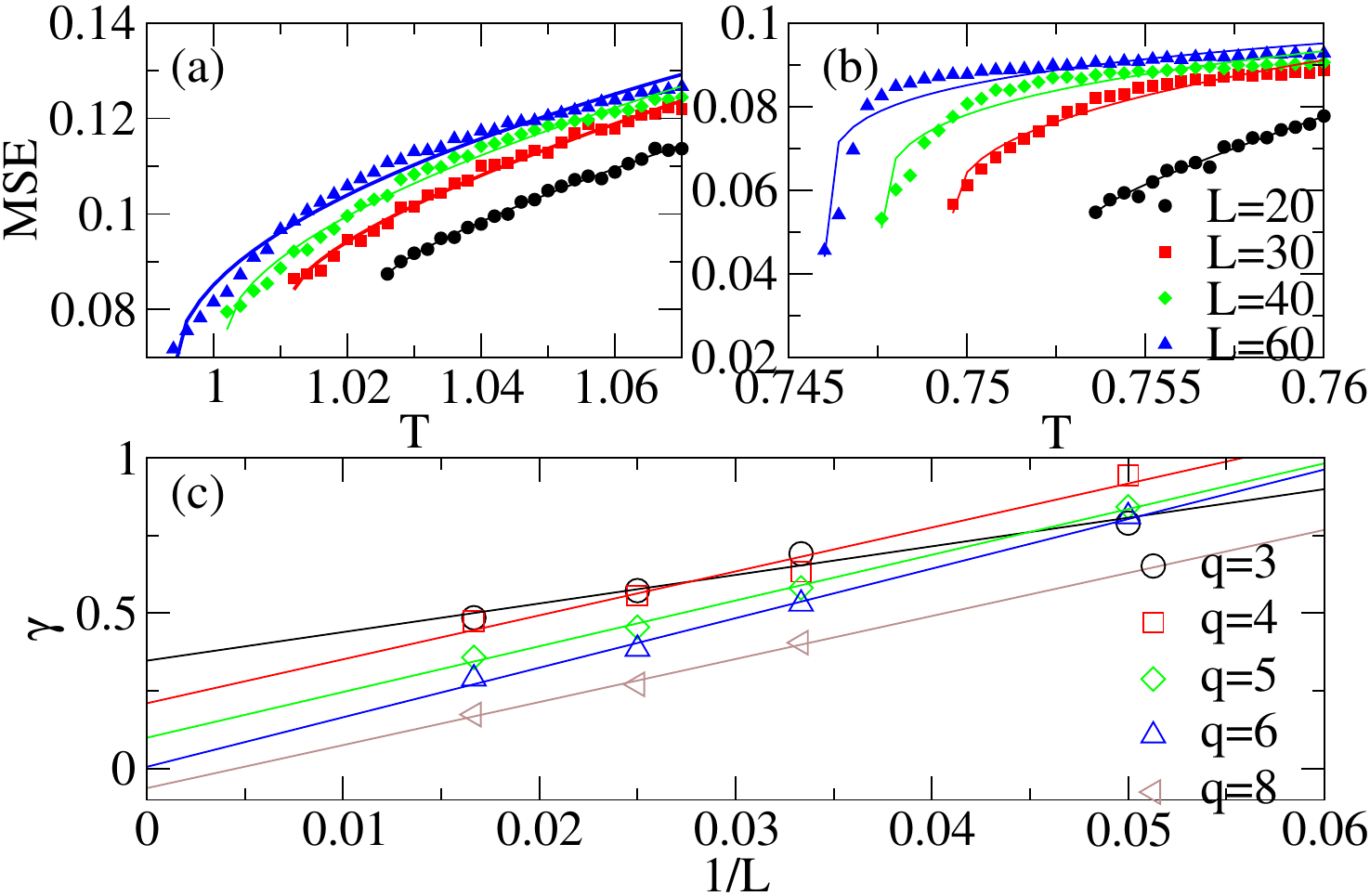}
\caption{Behaviors of the MSE around the phase transition points for (a) $q=3$ and (b) $q=8$ Potts models. The color lines represent the fitted curves as described by Eq.~\eqref{eqn:MSE_scaling}. (c) The fitted exponents $\gamma$ as functions of $1/L$ for various $q$'s.
For continuous transitions ($q\le4$), the values of $\gamma$ are positive; while $\gamma$'s become zero or negative for discontinuous transitions ($q>5$). For the exceptional case of $q=5$, where the transition is very weakly first-order, the fitted $\gamma$ is found to be slightly positive.
}%
\label{fig:potts_2}
\end{figure}
%------------------------- figure -------------------------------------

As a measure of the degree of disorder in the spin configurations, one expects that the MSE will change abruptly around the first-order transitions, while it may vary smoothly across the second-order ones. Such behaviors of the MSE are supported by our results, as seen by Figs.~\ref{fig:q2_potts}(a) and \ref{fig:q5_potts}(a) where the $q=2$ and the $q=5$ cases correspond to the second-order and the first-order transitions, respectively. Following the proposal in Ref.~\cite{Ho-Wang2023}, one may distinguish between these two kinds of transitions by examining the critical behaviors of the MSE around the phase transition points. We assume that the MSE takes its general form as the following scaling function,
\begin{equation}\label{eqn:MSE_scaling}
\mathrm{MSE}(T)=\mathrm{sgn}(T-T_c)\,A|T-T_c|^\gamma + B \;,
\end{equation}
where $\text{sgn}(T-T_c)$ is the sign function and $T_c$ is determined by the peak position of $L\Delta_\mathrm{MSE}$. By fitting our data for $T\gtrsim T_c$, the exponent $\gamma$ and the other two fitting parameters $A$ and $B$ can be obtained. For continuous transitions, one should have positive values of $\gamma$, while $\gamma\leq0$ implying a discontinuous jump in the MSE will be obtained for discontinuous transitions.

For illustration, fitting results of the MSE for several system sizes of the $q=3$ and the $q=8$ cases are shown in Figs.~\ref{fig:potts_2}(a) and (b). The scaling behaviors of the fitted exponents $\gamma$ for various values of $q$ are displayed in Fig.~\ref{fig:potts_2}(c). The estimated exponents $\gamma$ in the limit of $L=\infty$ are found to be positive for $q\leq4$, while the extrapolated findings with $\gamma\leq0$ are obtained for $q>5$. However, in the exceptional case of $q=5$, characterized by a very weakly first-order transition~\cite{1981JSP....24...69B}, the fitted exponent $\gamma$ shows a slightly positive value. It is worth noting that for this specific case, the analysis necessitates larger lattice sizes to account for the long correlation length. Nevertheless, excluding the case of $q=5$, the behavior of the MSE near transition temperatures can effectively discriminate between two distinct types of phase transitions.

The above results demonstrate the advantages of the present unsupervised learning technique in identifying and classifying phase transitions. Not only accurate values of the transition temperatures can be determined through the finite-size analysis of the peak positions of $L\Delta_\mathrm{MSE}$, the critical behaviors of the MSE around the phase transition points can be used to distinguish the discontinuous from the continuous transitions.

%%%%%%%%%%%%%%%%%%%%%%%%%%%%%%%%%%%%%%%%%%%%%%%%%%%%%%%%%%%%%%%%%%
\section{$q$-state clock model} \label{sec:q_clock}

To further explore the potential of the present machine learning technique in identifying phase transitions of distinct types, we consider in this section the $q$-state clock model on a square lattice~\cite{Lapilli_etal2006,%
Ortiz_etal2012,Kumano_etal2013,Li_etal2020,Li_etal2022,Chen_etal2022}. %
By changing the parameter $q$, both conventional continuous phase transitions of Landau-Ginzburg type and topological Berezinskii–Kosterlitz–Thouless (BKT) transitions~\cite{Berezinsky1970,Kosterlitz_Thouless1973,Kosterlitz1974} can appear. The phase transitions discussed in Sec.~\ref{sec:q_Potts} belong to the Landau-Ginzburg type. The BKT phase transition is associated with the unbinding of vortex-antivortex pairs and cannot be characterized by spontaneous symmetry breaking with local order parameter. Identifying such a transition proves challenging and innovative methods are required to pinpoint it. We show below that the modified anomaly detection with CAE can locate the BKT transitions with success.

The Hamiltonian of the $q$-state clock model, also known as the planar Potts model~\cite{Potts_review}, is
\begin{equation}\label{eqn:q_clock}
H_\mathrm{clock}=-\sum_{\langle i,j\rangle}\cos(\theta_i- \theta_j) \;,
\end{equation}
where the $q$-state spin on site $i$ is denoted by a planar angle of spin orientation $\theta_i=2\pi n_i/q$ with $n_i=0,1,\cdots,q-1$, and $\langle i,j\rangle$ stands for the nearest neighbors.

The $q$-state clock model is exactly solvable for $q\leq4$ and their phase transitions are of Landau-Ginzburg type driven by fluctuating local order parameters with symmetry breaking.
For $q>4$, besides the low-temperature ferromagnetic ordered phase and the high-temperature paramagnetic disordered phase, an intermediate quasi-long-range ordered BKT phase emerges. Thus, there are two distinct BKT transitions driven by topological defects (vortices) at temperatures $T_{c1}$ and $T_{c2}\;(>T_{c1})$. In the limit of $q\to\infty$, the model is equivalent to the standard 2D $XY$ model, in which the BKT phase extends throughout the low-temperature regime, and thus gives $T_{c1}=0$.

%------------------------- figure -------------------------------------
\begin{figure}[htb]
\includegraphics[width=\linewidth]{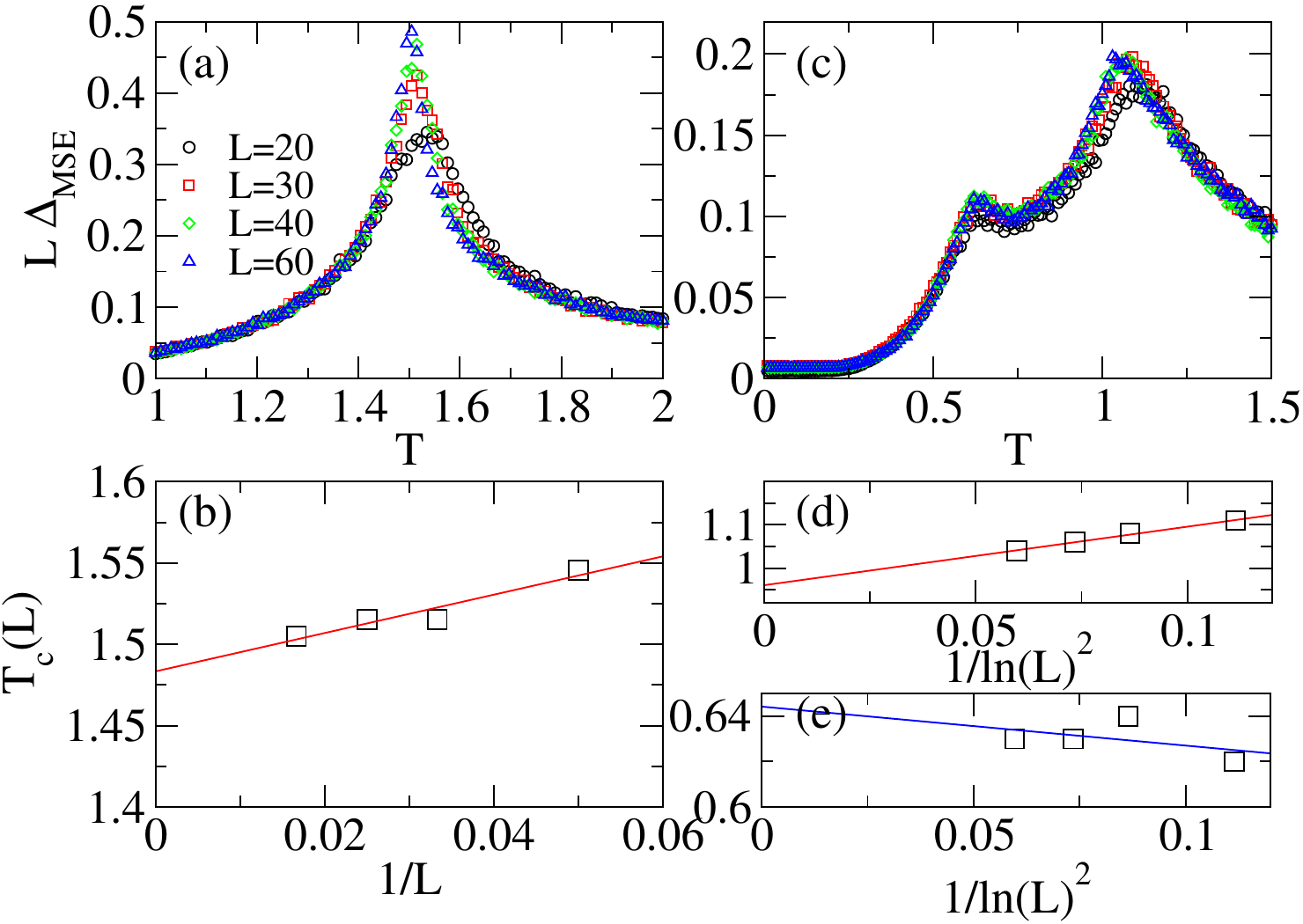}
\caption{The determination of critical temperatures for the (a)-(b) $q=3$ and (c)-(e) $q=6$ clock models. (a) The $L\Delta_\mathrm{MSE}$ of $q=3$ model exhibits a single pronounced peak at the transition. (b) The finite-size analysis for the critical temperature $T_c$ gives the extrapolated value $T_c=1.483$. (c) For the case of $q=6$, two BKT transitions can be clearly identified as local peaks. The extrapolated values of the critical temperatures are found to be $T_{c1}=0.644$ (e) and $T_{c2}=0.9605$ (d), in agreement with previous findings.}
\label{fig:clock}
\end{figure}
%------------------------- figure -------------------------------------

Following the discussions in Sec.~\ref{sec:q_Potts}, the results of the rescaled standard deviation $L\Delta_\mathrm{MSE}$ for the cases of $q=3$ and $q=6$ are presented in Fig.~\ref{fig:clock}. The peak positions of $L\Delta_\mathrm{MSE}$ correspond to the critical temperatures for given system sizes. Notice that we use here the two-component spin vector $(\cos\theta_i, \sin\theta_i)$ on each site $i$ as input and output data, and each configuration thus has $2\times L^2$ elements. As a result, the input/output layers of the CAE in Fig.~\ref{fig:CAE} consist of two layers in the present case.

For $q=3$, the model is equivalent to the three-state Potts model, which has a single critical temperature $T_c=3/[2\ln(1+\sqrt{3})]\simeq1.492$~\cite{Ortiz_etal2012}. As seen from Figs.~\ref{fig:clock}(a) and (b), $L\Delta_\mathrm{MSE}$ in this case does show a single pronounced peak and the extrapolated critical temperature $T_c=1.483$ agrees well with the theoretical prediction. In contrast, Fig.~\ref{fig:clock}(c) shows that $L\Delta_\mathrm{MSE}$ for the $q=6$ case exhibits a double-peak structure, indicating two phase transitions as expected. Since the transitions are anticipated to be of the BKT type, we employ the finite-size scaling analysis based on the Kosterlitz’s expression for the temperature dependence of the correlation length~\cite{Kosterlitz1974}. The scaling function is thus given by
\begin{equation}\label{BKT_scaling}
T_c(L)=T_c+\frac{b}{[\ln(L)]^2}\;,
\end{equation}
where $b$ is a fitting parameter. From
Figs.~\ref{fig:clock}(d) and (e), we obtain the extrapolated values of the low-temperature and the high-temperature critical points as $T_{c1}=0.644$ and $T_{c2}=0.9605$, respectively. Our findings are comparable with previous known results (see Table I in Ref.~\cite{Li_etal2022}). More precise values could be reached if data of larger system sizes are included to get rid of finite-size effects.

The above results clearly demonstrate that the present machine learning technique can recognize not only the conventional order-disorder phase transitions but also the topological transitions of the BKT type. Notice that our approach only imports the spin configurations. Despite this, it can still differentiate the intermediate phase with quasi-long-range order from the ordered and the disordered ones. Furthermore, by adopting finite-size scaling analysis, the critical temperatures can be determined with high accuracy.

%%%%%%%%%%%%%%%%%%%%%%%%%%%%%%%%%%%%%%%%%%%%%%%%%%%%%%%%%%%%%%%%%%
\section{Generalized XY model} \label{sec:GXYmodel}

Encouraged by the success we've achieved, we now proceed to examine the generalized XY (GXY) model~\cite{Korshunov1985,Lee-Grinstein1985,%
D_B_Carpenter_1989,Poderoso_etal2011,Dian-Hlubina2011,Hubscher-Wessel2013,%
Canova_etal2014,Canova_etal2016,Nui_etal2018,Song-Zhang2021} %
on a square lattice to explore the broader utility of our machine learning approach. Due to the competition between two interaction terms, this model permit a much richer variety of spin configurations, consequently leading to more intricate phase diagrams and a greater diversity of phase transitions.

The Hamiltonian of the GXY model is given by
\begin{equation}\label{eqn:GXY}
H_\mathrm{GXY}=-\sum_{\left\langle i,j \right\rangle} \{\Delta \cos( \theta_i - \theta_j)+ (1-\Delta) \cos[q(\theta_i-\theta_j)]\} \; ,
\end{equation}
where $\langle i,j\rangle$ denotes nearest neighbor sites and $\theta_i\in(0, 2\pi]$ is the planar angle of spin orientation at site $i$. Besides, $\Delta$ gives the relative weight of the pure XY model and $q$ is an integer parameter. The second term in Eq.~\eqref{eqn:GXY} describes a competing interaction with $2\pi/q$ period, which could drives the system to form a generalized nematic phase with $q$ preferred spin orientations. We focus on the $q=2$ case in the present study.

%---------------------------------------
\begin{figure}[htb]
\includegraphics[width=\linewidth]{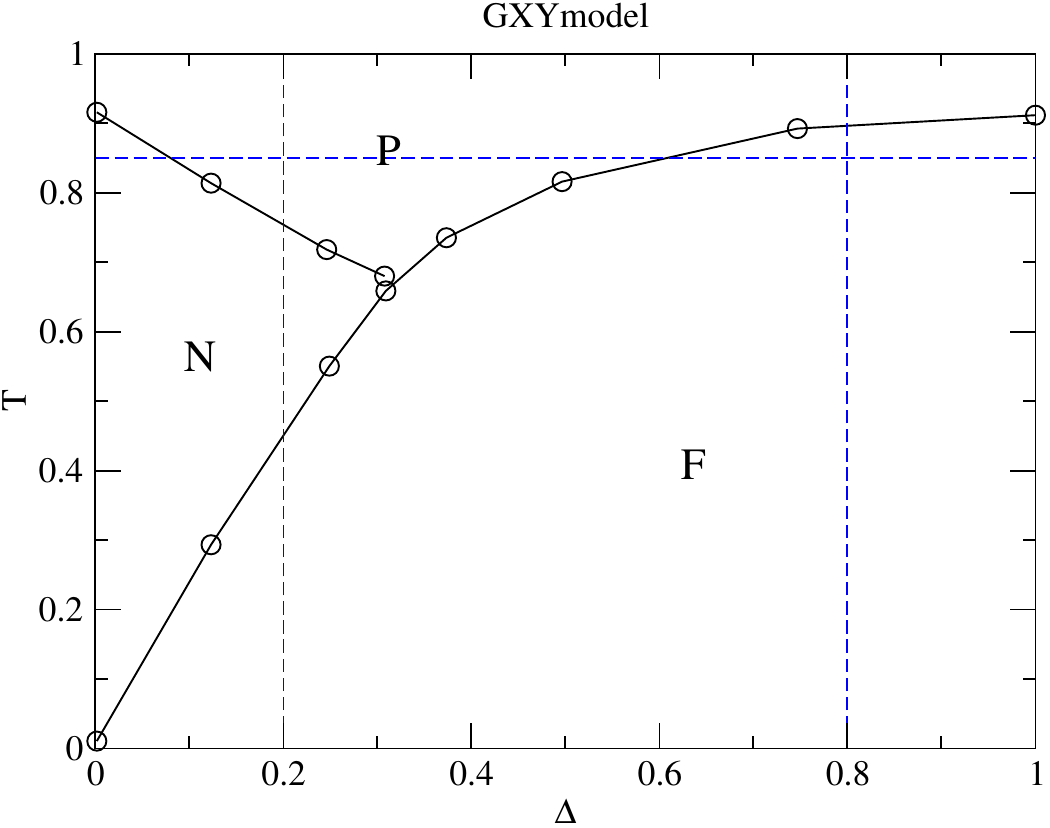}
\caption{Phase diagram of the GXY model with $q=2$, where the data come from Ref.~\cite{D_B_Carpenter_1989}. The symbols $N$, $F$, $P$ represent the nematic, the ferromagnetic and the paramagnetic phases, respectively. Both of the $N$-$P$ and the $F$-$P$ transitions belong to the BKT universality class, while the $N$-$F$ transition is of the Ising type. The dashed lines indicate the parameter paths to be scanned in Figs.~\ref{fig:gXY_1} and \ref{fig:gXY_2}}	
\label{fig:gXY_phase}
\end{figure}
%---------------------------------------

The $q=2$ GXY model has a rich phase diagram, as shown in Fig.~\ref{fig:gXY_phase}. For both cases of $\Delta=0$ and 1, the model reduces to the pure XY model (redefining $2\theta\to\theta$ in the first case) and thus has the same transition temperature as that of the pure XY model. When $\Delta=0$, the system at low temperatures belongs to the quasi-long-range nematic ($N$) phase consisting of half-integer vortices connected by strings (domain walls)~\cite{Lee-Grinstein1985}. In contrast, the system with $\Delta=1$ is an integer vortex binding  phase in low-temperature limit and carries quasi-long-range ferromagnetic ($F$) order. At higher temperatures, the system becomes disordered and is in a paramagnetic ($P$) phase. Both of the $N$-$P$ and the $F$-$P$ phase transitions belongs to the BKT universality class. Because the $N$-$P$ transition is associated with the unbinding of half-integer vortices and antivortices, it is often referred to as a half BKT transition. On the other hand, the transition from the $F$ to the $N$ phases, induced by the proliferation of domain walls linking half-integer vortices, belongs to the Ising universality class.

%---------------------------------------
\begin{figure}[htb]
\includegraphics[width=\linewidth]{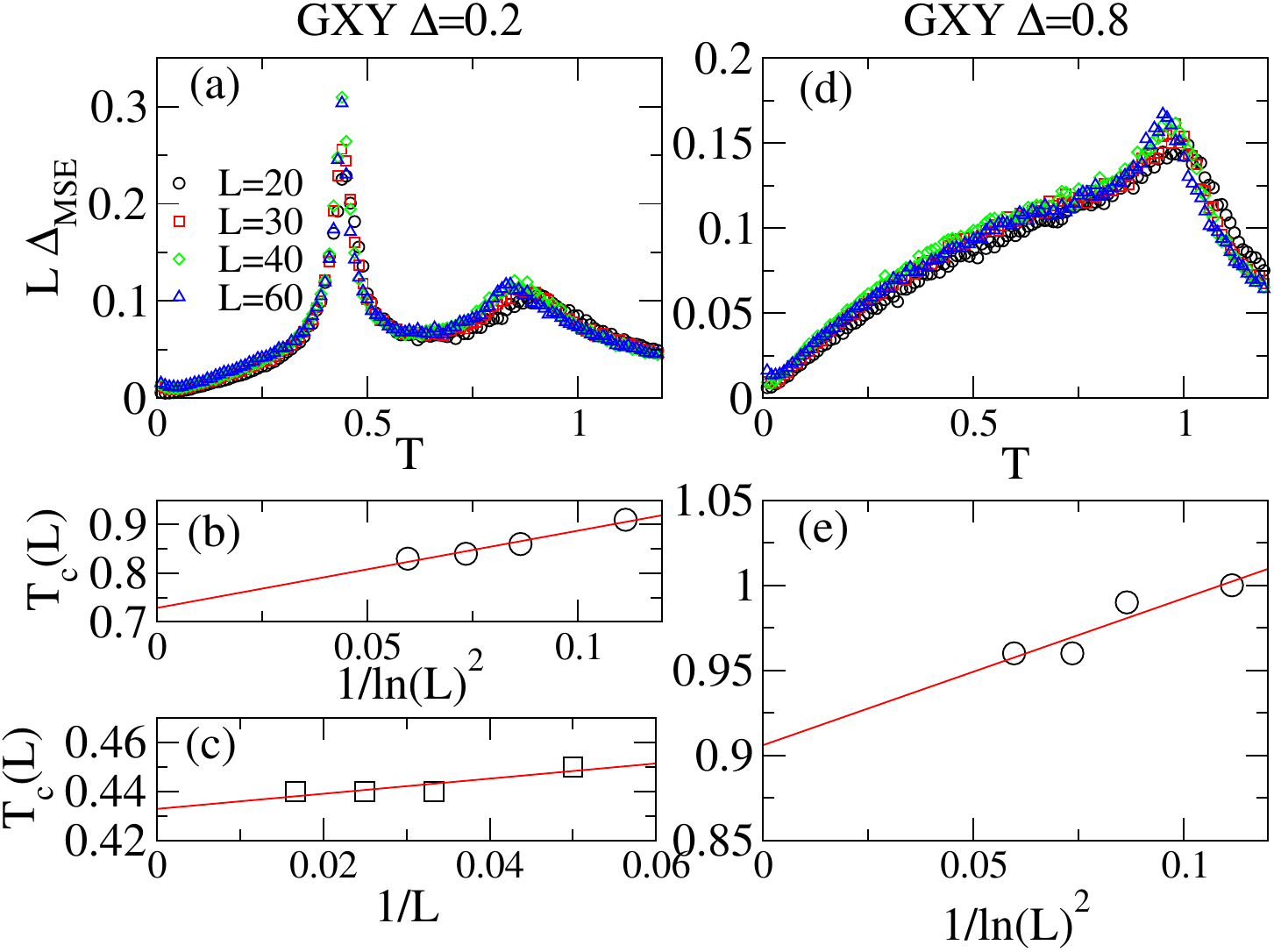}
\caption{Results for the GXY model with $q=2$ for (a)-(c) $\Delta=0.2$ and (d)-(e) $\Delta=0.8$. In (a), both the $F$-$N$ and the $N$-$P$ transitions can be clearly identified by the peaks in the rescaled standard deviation $L\Delta_\mathrm{MSE}$. The extrapolated values of the critical temperatures are found to be $T_{c1}=0.433$ (c) and $T_{c2}=0.729$ (b), respectively. (d)  $L\Delta_\mathrm{MSE}$ for $\Delta=0.8$ showing a single $F$-$P$ transition.  (e) The finite-size analysis gives the extrapolated critical temperature  $T_c=0.906$.}
\label{fig:gXY_1}
\end{figure}
%---------------------------------------

We now apply our machine learning technique to determine the transition temperatures in the GXY model. To distinguish among the three phases and learn all the transitions, we need to utilize both configurations of the spin vector $(\cos\theta_i, \sin\theta_i)$ and the nematic director $(\cos2\theta_i, \sin2\theta_i)$ on each site $i$ as input and output data. This means that each site has four elements. Consequently, in the present case, the CAE illustrated in Fig.~\ref{fig:CAE} comprises four layers in its input/output design.

The results of the rescaled standard deviation $L\Delta_\mathrm{MSE}$ for the cases of $\Delta=0.2$ and 0.8 are presented in Fig.~\ref{fig:gXY_1}. The peak positions of $L\Delta_\mathrm{MSE}$ correspond to the critical temperatures for given system sizes. Fig.~\ref{fig:gXY_1}(a) shows that $L\Delta_\mathrm{MSE}$ for the $\Delta=0.2$ case exhibits a double-peak structure, indicating two phase transitions as expected. Given that the $N$-$P$ transition takes place at a higher critical temperature $T_{c2}$ and follows a BKT-type behavior, we employ the scaling function in Eq.~\eqref{BKT_scaling} to compute its critical temperature in the thermodynamic limit. Conversely, for the $F$-$N$ transition occurring at a lower critical temperature $T_{c1}$ and being of Ising type, we utilize the conventional finite-size scaling method. As shown by Figs.~\ref{fig:gXY_1}(b) and (c), the extrapolated values are $T_{c1}=0.433$ and $T_{c2}=0.729$. These values agree well with recent results $T_{c1}=0.436$ and $T_{c2}=0.727$ obtained by large-scale Monte Carlo simulations~\cite{Nui_etal2018}. In contrast, for $\Delta=0.8$, Figs.~\ref{fig:gXY_1}(d) and (e) show a single pronounced peak and the extrapolated critical temperature $T_c=0.906$ by using the BKT scaling function, respectively. Our value is again in agreement with that ($T_c=0.885$) found by  Monte Carlo simulations~\cite{Nui_etal2018}.

%---------------------------------------
\begin{figure}[htb]
\includegraphics[width=\linewidth]{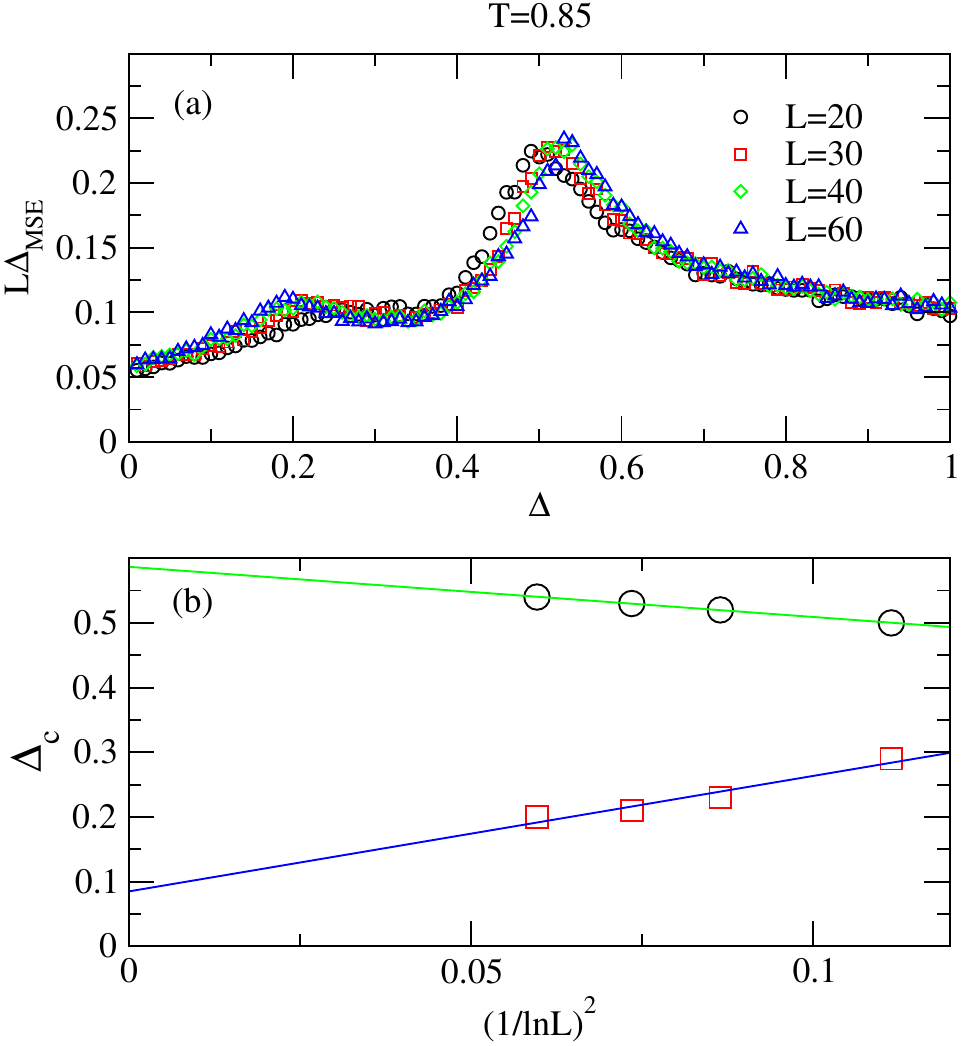}
\caption{Results for the GXY model with $q=2$ for various $\Delta$ at a fixed temperature $T=0.85$. (a) Both the $N$-$P$ and the $P$-$F$ transitions are identified by the peaks of the rescaled standard deviation $L\Delta_\mathrm{MSE}$. (b) The extrapolated values of critical $\Delta_c$ are determined to be 0.085 and 0.587, respectively. }
\label{fig:gXY_2}
\end{figure}
%---------------------------------------

As a further test, we consider the case by varying $\Delta$ at a fixed temperature $T=0.85$. The results of the rescaled standard deviation $L\Delta_\mathrm{MSE}$ and the finite-size analysis for the critical temperatures are presented in Fig.~\ref{fig:gXY_2}. 
Again, the training set includes data from ten evenly separated $\Delta$ within the parameter region.
 Two peaks in $L\Delta_\mathrm{MSE}$ are observed, which corresponds two distinct BKT transitions. By using the BKT scaling function in Eq.~\eqref{BKT_scaling}, the extrapolated values of critical $\Delta_c$ are determined to be 0.085 and 0.587, respectively. These values are consistent with those obtained by recent tensor network calculations, $\Delta_{c}=0.10$ and 0.58~\cite{Song-Zhang2021}.

Our calculations reveal that, by using both the spin and the nematic configurations as input data, delicate difference between two distinct quasi-long-range ordered phases, specifically the $F$ and the $N$ phases, can be effectively detected by the present machine learning approach. In addition, the location of the corresponding phase boundaries can be accurately determined under finite-size scaling analysis. Furthermore, two distinct BKT transitions involving unbinding of either integer or half-integer vortices are unambiguously discriminated by our approach. Motivated by the success in the present studies, we believe our approach can be extended to address more general cases.

%%%%%%%%%%%%%%%%%%%%%%%%%%%%%%%%%%%%%%%%%%%%%%%%%%%%%%%%%%%%%%%%%%
\section{discussions and conclusions} \label{sec:conclusion}

In this investigation, we employ an unsupervised machine learning technique in conjunction with finite-size scaling analysis to ascertain the critical temperatures of various classical spin models. Instead of using correlation functions and/or order parameters in some other approaches, the in-situ spin configurations without further processing are taken in our scheme as the input data for learning.
We find that, by training with input data across from all phases, the MSE of CAE neural networks can be regarded as a metric for quantifying the degree of disorder to characterize different phases.
As a result, the rescaled standard deviation of the MSE proves to be a proper indicator for identifying phase boundaries, which displays peaks at phase transition points. Importantly, this behavior remains true regardless of the specific structure or the hyperparameters of the CAE, as long as the CAE possesses the capability to capture structural information from input states characterizing the inherent order of the phase. Therefore, unlike other applications of CAEs, our approach does not necessitate the additional effort of minimizing the MSE. In our case, the training process is highly efficient, with a modest number of training epochs (only 20). Our CAE model is characterized by its simplicity and efficiency, as it encodes all essential phase information in the MSE and its distribution, eliminating the need for determining the number of phases independently \cite{Tasi_etal2021,Chung_etal2023,Kottmann_etal2021} or performing additional computations \cite{Acevedo_etal2021} to extract critical points.  

In conclusion, our approach has successfully and accurately identified the locations of critical points across various statistical models, including those featuring topological phases. Furthermore, it enables the determination of transition types through an analysis of the critical behavior of the MSE. The straightforward and efficient nature of our approach make it easy to extend to other statistical models, and we anticipate to explore its potential application in other contexts, such as other topological states, frustrated states, and even quantum phases in future work. 
As only the in-situ spin configurations are employed as the input data for learning, our scheme can be applied as well to the cases with experimentally accessible data as inputs.

%%%%%%%%%%%%%%%%%%%%%%%%%%%%%%%%%%%%%%%%%%%%%%%%%%%%%%%%%%%%%%%%%%
\begin{acknowledgments}
The authors would like to thank Ching-Yu Huang for enlightening disccussions. This research was supported by Grant No. NSTC 112-2112-M-029-005 of the National Science and Technology Council of Taiwan. M.F.Y. and K.K.N also acknowledge the supports from the National Science and Technology Council of Taiwan under Grant No. MOST 111-2112-M-029-004 and Grant No. MOST 111-2112-M-029-008, respectively.
\end{acknowledgments}
%%%%%%%%%%%%%%%%%%%%%%%%%%%%%%%%%%%%%%%%%%%%%%%%%%%%%%%%%%%%%%%%%%

\bibliography{refs,refs_fidelity,refs_models}

\end{document}